\documentclass[final,5p,times,twocolumn]{elsarticle}
\usepackage{amssymb,amsmath,amsthm,bm,graphicx,epsfig,lineno,color,hyperref}
\usepackage{textcomp}
\usepackage{pifont}
\usepackage{bbding}
\usepackage{latexsym}
\usepackage{amsfonts}
\usepackage{float}
\usepackage{lipsum}
\usepackage{txfonts}
\usepackage{yfonts}
\usepackage{soul}

\usepackage{hyperref}

\usepackage{amsthm}
\theoremstyle{definition}
\newtheorem{remark}{Remark}[section]

\usepackage[OT2,OT1]{fontenc}
\newcommand\cyr{%
\renewcommand\rmdefault{wncyr}%
\renewcommand\sfdefault{wncyss}%
\renewcommand\encodingdefault{OT2}%
\normalfont
\selectfont}
\DeclareTextFontCommand{\textcyr}{\cyr}

\journal{Mech.\ Res.\ Commun.\   \hspace{1.31in} {\rm \normalsize 1}}

\newcommand{\be}{\begin{equation}}
\newcommand{\en}{\end{equation}}

\newcommand{\lshad}{[\![}
\newcommand{\rshad}{]\!]}
\newcommand{\rd}{\mathrm{d}}

\newcommand{\ri}{\mathrm{i}}

\DeclareMathOperator{\sinc}{sinc}

\begin{document}

\begin{frontmatter}

\title{Acoustic shock and acceleration waves in selected  inhomogeneous fluids}

\author[NRL]{R.S.\ Keiffer} 
\author[NRL]{P.M.\ Jordan\corref{cor1}}
\cortext[cor1]{Corresponding author.}
\address[NRL]{Acoustics Div., U.S.\ Naval Research Laboratory, Stennis Space Ctr.,  
 Mississippi 39529, USA} 

\author[PU]{I.C.\ Christov}
\address[PU]{School of Mechanical Engineering, Purdue University, West Lafayette, Indiana 47907, USA}

\begin{abstract}

Acoustic  shock and acceleration waves in inhomogeneous fluids are investigated using both analytical and numerical methods.  In the context of  start-up signaling problems,   and based on linear acoustics theory,  we study the propagation of such waveforms  in the atmosphere and in fluids that possess a periodic ambient density profile. It is shown that vertically-running shock and acceleration waves in the atmosphere suffer amplitude growth. In contrast, those in the periodic-density  fluid have bounded amplitudes that exhibit  periodic, but non-trivial, oscillations; this is illustrated via a series of numerically-generated profile-evolution plots, which were  computed using the PyClaw software package.

\end{abstract}


\end{frontmatter}



\section{Introduction}\label{sect:Intro}

The central focus of this paper is   singular surfaces, specifically, acoustic shock and acceleration waves,  in inhomogeneous fluids.   The study of acoustic  propagation in such media dates back more than two centuries to the works of Laplace and Poisson on propagation in the atmosphere; see, e.g., Refs.~\cite[\S 303]{Lamb06}, \cite[p.~553]{Lamb11}, and those cited therein.  The modern treatment of acoustic phenomena in inhomogeneous fluids, however, can be traced back to the early 20th century and the works of Lamb~\cite{Lamb06,Lamb09,Lamb11}, who, like his forerunners\footnote{Meaning, of course,  Laplace and Poisson, but also Rayleigh~\cite{Rayleigh}.}, considered propagation in the atmosphere. The problem of acoustic waves in inhomogeneous media~\cite{OW} remains of significant interest due, e.g.,  to the importance of sound interaction with both the atmospheric boundary layer and features in the 
terrain~\cite{WPO}. Yet, most of the current work remains purely computational.

On the other hand, the results of the singular surface analyses carried out below are exact. The three cases considered are based on linear acoustics theory and involve the simplest density profiles used to model inhomogeneous fluids.  Our aims  are to shed light on how shock and acceleration waves evolve in such fluids and highlight the effectiveness of singular surface theory as a tool to probe inhomogeneous media in general.  

To this end, the present article is organized as follows. In Sect.~\ref{sect:Sys_Gen},  the system of Euler equations governing compressible flow in inhomogeneous fluids is stated and  terms/quantities are defined.
In Sect.~\ref{sect:Atmos}, we revisit the issue of vertical  propagation in the atmosphere, extending a number of Lamb's studies~\cite{Lamb11,Lamb45}  to include shock and acceleration waves.  Then, in Sect.~\ref{sect:Periodic}, we do the same for the case of a fluid that exhibits a periodic ambient density profile and is free of external body forces, and we also provide numerical results and details of the implementation of our model into a modern shock-capturing numerical software package. Finally, in Sect.~\ref{sect:Closure} we note, and briefly discuss, three possible extensions of the present investigation.

Before stating our governing system, however, it should be noted that researchers in continuum physics have investigated  singular surfaces not only in fluids, but also in  solids; see, e.g., Refs.~\cite{BH68,chen1973,S11,tt60} and those cited therein.  Indeed,  the present study was inspired, in part, by the work of Berezovski et al.~\cite{BEM00,BM02} on waves in inhomogeneous thermoelastic media, which extended Maugin's~\cite{M93} work on material inhomogeneties in elasticity to thermoelastic media. 
And as we do here, Berezovski et al.\ employed singular surface theory and presented numerical simulations using LeVeque's~\cite{L97} shock-capturing scheme; see also Maugin's~\cite{M98} discussion of shock waves, singular surfaces, and phase-transition fronts.

\section{Euler equations for compressible flow in inhomogeneous fluids}\label{sect:Sys_Gen}

In the case of an inhomogeneous fluid, which we assume to be lossless\footnote{That is, the flow is 
\emph{isentropic}~\citep[p.~60]{T72}; see Eq.~\eqref{eq:EoS_1}.},  the  (Euler) system of equations that governs  compressible flow  becomes~\cite{Bergmann46}:
\begin{subequations}\label{sys:Comp}
\begin{align}
D \varrho/Dt &= -\varrho (\boldsymbol{\nabla} \boldsymbol{\cdot} {\bf u}),\label{eq:cont_1}\\
\varrho D{\bf u}/Dt  &= -\nabla p + \varrho {\bf b}, \label{eq:mom_1}\\
D p/Dt &= c^2 D\varrho/Dt \qquad (D\eta/Dt = 0).\label{eq:EoS_1}
\end{align}
\end{subequations}
Here, ${\bf u}=(u,\upsilon,w)$ is the velocity vector; $\varrho (>0)$ is the mass density; $p (>0)$ is the thermodynamic pressure; $\eta$ is the specific entropy; ${\bf b}={\bf b}(x,y,z)$ is the external (per unit mass) body force  vector;    $D/Dt$ is the material derivative; and the (thermodynamic) variable $c(>0)$ denotes the sound speed. 

For fluids in general, $c^2 = A/\varrho$, where $A$ is the adiabatic bulk modulus~\cite[p.~30]{P89}.  In the case of \emph{perfect gases}\footnote{Also known as \emph{polytropic gases}~\cite[p.~154]{W74}; see Thompson~\cite[\S 2.5]{T72}.}, however, $A = \gamma p$;  therefore, in such gases, $c^2 = \gamma p/\varrho$ and, moreover, $p$, $\varrho$, and $\vartheta$ satisfy  the following special case of the \emph{ideal gas law}:
\be\label{eq:EoS_ideal}
p=(c_p-c_v)\varrho \vartheta \qquad (c_{p}, c_{v} := \textrm{const.}).
\en
Here, $\vartheta(>0)$ is the absolute temperature; $c_p > c_v >0$ are the specific heats at constant pressure and volume, respectively; and $\gamma = c_p/c_v$,  where  $\gamma \in (1, 5/3]$ for perfect gases.

In what follows, we shall  restrict our attention to propagation in 1D and investigate linearized versions of Sys.~\eqref{sys:Comp}. Moreover, the ambient state of the fluid shall always be taken as  \emph{quiescent}~\cite[p.~14]{P89}; i.e., while $p_{\rm a}$,  $\varrho_{\rm a}$,  $\vartheta_{\rm a}$, and $\eta_{\rm a}$ may vary with, at most, position,  ${\bf u}_{\rm a}  = (0,0,0)$, where  a subscript `a'   denotes the ambient state value of the quantity to which it is attached. And lastly,  we  reserve  `$\zeta$'  for use  hereafter as a `dummy' variable.


\section{Vertical propagation in the atmosphere}~\label{sect:Atmos}

In this section  we consider the  special case of Sys.~\eqref{sys:Comp} discussed in Ref.~\cite[p.~159]{W74}, wherein  ${\bf u}=(0,0,w(z,t))$, $p=p(z,t)$, $\varrho =\varrho(z,t)$,  and ${\bf b} = (0,0,-g)$; here, $g$ denotes the acceleration due to gravity near the surface, where $g \approx 9.81\textrm{m/s}^2$ in the case of Earth, and the $+z$-axis is directed vertically upwards. 

And for later reference we observe that, for perfect gases,
\be\label{eq:entropy_gas}
\eta_{\rm a}(z)-\eta_{0} = c_{v}\Big\{ \ln[p_{\rm a}(z)/p_{0}]-\gamma \ln [\varrho_{\rm a}(z)/\varrho_{0}]\,\Big \},
\en
a relation  easily derived from, e.g.,  Ref.~\cite[Eq.~(19)]{PMJ16}.  Here, we introduce the  notation $p_{0} := \lim_{z\to 0}p_{\rm a}(z)$, $\varrho_{0} := \lim_{z\to 0}\varrho_{\rm a}(z)$,    $\vartheta_{0} := \lim_{z\to 0}\vartheta_{\rm a}(z)$, and $\eta_{0} := \lim_{z\to 0}\eta_{\rm a}(z)$, where $p_{0}$, $\varrho_{0}$,    $\vartheta_{0}$, and $\eta_{0}$  are constants, with $p_{0}$, $\varrho_{0}$,  and  $\vartheta_{0}$ connected via Eq.~\eqref{eq:EoS_ideal}.

\subsection{Linearized system and equation of motion}

We begin by eliminating $D\varrho/Dt$ between Eqs.~\eqref{eq:cont_1} and~\eqref{eq:EoS_1}.  Then, on setting  $\tilde{p}(z,t) = p(z,t) -p_{\rm a}(z)$ and $\tilde{\varrho}(z,t) = \varrho(z,t) -\varrho_{\rm a}(z)$ and  linearizing about the ambient state,  Sys.~\eqref{sys:Comp}  becomes
\begin{subequations} \label{sys:atmos-1}\begin{align}
	\tilde{\varrho}_{t}+\varrho_{\rm a}^{\prime}(z) w +\varrho_{\rm a}(z) w_{z} &= 0,\label{eq:cont_atmos-1}\\
	\varrho_{\rm a}(z)w_{t} + \tilde{p}_{z} &= - \{p_{\rm a}^{\prime}(z)+g[\tilde{\varrho}+\varrho_{\rm a}(z)]\},\label{eq:mom_atmos-1}\\
	\tilde{p}_{t}+A_{\rm a}(z)w_{z} &= -p_{\rm a}^{\prime}(z)w,\label{eq:EoS_atmos-1}
\end{align}\end{subequations} 
 where in this section a prime denotes $\rd /\rd z$.

Making use of the `statical' relation (Lamb~\cite[p.~541]{Lamb45})
\be\label{eq:statical_rel}
p_{\rm a}^{\prime}(z) =-g\varrho_{\rm a}(z), 
\en
which stems  from the fact that the field variables' ambient state values  must satisfy Sys.~\eqref{sys:atmos-1}, yields the further simplification
\begin{subequations}\label{sys:atmos-2}\begin{align}
\tilde{\varrho}_{t}+\varrho_{\rm a}^{\prime}(z) w +\varrho_{\rm a}(z) w_{z} &= 0,\label{eq:cont_atmos-2}\\
\varrho_{\rm a}(z)w_{t} + \tilde{p}_{z} &= - g\tilde{\varrho},\label{eq:mom_atmos-2}\\
\tilde{p}_{t}+A_{\rm a}(z)w_{z} &= g\varrho_{\rm a}(z)w.\label{eq:EoS_atmos-2}
\end{align}\end{subequations} 
Now  eliminating $\tilde{p}_{t}$ between Eqs.~\eqref{eq:mom_atmos-2} and~Eq.~\eqref{eq:EoS_atmos-2}, after applying  $\partial/\partial t$ to the former and $\partial/\partial z$ to the latter, yields
\be\label{eq:mom_atmos-w}
\varrho_{\rm a}(z)w_{tt}+ [-A_{\rm a}(z) w_{z}]_{z}=0,
\en
where we have also made use of Eq.~\eqref{eq:cont_atmos-2}.  On setting $\wp_{t}(z,t) :=-A_{\rm a}(z) w_{z}$ in Eq.~\eqref{eq:mom_atmos-w} and then integrating with respect to $t$, we are led to consider the \emph{two-equation} system
\begin{subequations}\label{Sys:atmos_wp}\begin{align}
	\varrho_{\rm a}(z)w_{t} + \wp_z &= 0,\\
	\wp_t + A_{\rm a}(z) w_{z} &= 0,
\end{align}\end{subequations}
where the resulting function of integration has been set equal to zero without loss of generality.

In anticipation of the singular surface analyses that shall be carried out in the next subsection, let us divide Sys.~\eqref{Sys:atmos_wp} by $\varrho_{\rm a}(z)$, and set $c_{\rm a}^{2}(z) :=A_{\rm a}(z)/\varrho_{\rm a}(z)$. After further simplifying and then setting $\psi(z,t) :=\wp(z,t)/\varrho_{\rm a}(z)$,  Sys.~\eqref{Sys:atmos_wp} becomes
\begin{subequations} \label{Sys:atmos_psi}
\begin{align}
	w_{t} + \psi_{z} &= -[\varrho_{\rm a}'(z)/ \varrho_{\rm a}(z)] \psi, \label{eq:flux_psi}\\
	\psi_{t}+ c_{\rm a}^{2}(z) w_{z} &= 0.\label{eq:conv_psi}
\end{align}
\end{subequations}


It is a straightforward matter to now eliminate $\psi$  between the PDEs of Sys.~\eqref{Sys:atmos_psi} and obtain the equation of motion governing  vertical propagation in an atmosphere (and ocean), namely,
\be\label{eq:EoM_atmos_fluid}
w_{tt} =  c_{\rm a}^{2}(z)w_{zz} +[A_{\rm a}^{\prime}(z)/\varrho_{\rm a}(z)]w_{z}.
\en
In the case of a perfect gas, wherein $A_{\rm a}(z) = \gamma p_{\rm a}(z)$, Eq.~\eqref{eq:EoM_atmos_fluid}, with the aid of Eq.~\eqref{eq:statical_rel}, reduces to
\be\label{eq:EoM_atmos_gas}
w_{tt} = c_{\rm a}^{2}(z)w_{zz} -\gamma g w_{z}.
\en
Equations~\eqref{eq:EoM_atmos_fluid} and~\eqref{eq:EoM_atmos_gas} are, we observe,  equivalent to the first and second displayed PDEs in Ref.~\cite[p.~160]{W74}. Notice also that both Eqs.~\eqref{eq:EoM_atmos_fluid} and \eqref{eq:EoM_atmos_gas} are second-order wave equations with variable coefficients. A  survey of such PDEs is given 
in~\cite{CC17}, wherein it is shown that they also arise  in propagation problems involving \emph{homogeneous} media with moving boundaries.

In this section we shall, for the two most common cases of $\varrho_{\rm a}(z)$ (relating to the atmosphere), investigate the following \emph{hybrid}\footnote{Time- and Laplace-domain BCs are used  to select the solution that is \emph{initially}, i.e., prior to encountering any boundary that might be present, right-propagating. The idea for this hybrid formulation came from the problem treated in Ref.~\cite[\S 82]{CJ63}, which also exemplifies the fact that the spatial asymptotic behavior of a  Laplace transform need not be reflected in its inverse.} initial-boundary value problem (hIBVP):
\begin{subequations}\label{IBVP:Atmos_Gen}
\begin{align}
\label{IBVP:Atmos_Gen_EoM}
& w_{tt} =  c_{\rm a}^{2}(z)w_{zz} -\gamma g w_{z}, \quad (z,t)\in (0, \lambda)\times (0, \infty),\\
& w(0,t) = W_{0}\Theta (t)f(t),\quad |\overline{w}(\lambda,s)| < \infty, \quad  t, s>0,\\
& w(z,0) = 0, \quad  w_{t}(z,0) = 0, \quad  z \in (0, \lambda),
\end{align}
\end{subequations}
where $W_0(> 0)$ is a  constant, $s$ is the Laplace transform parameter, a bar over a quantity denotes the image of that quantity in the Laplace transform domain,   
and $\lambda (>0)$ will either be assigned a (fixed) value or be replaced with $\infty$.  Also, in this communication we let
\be
f(t) :=\begin{cases}
1 & \textrm{$\Rightarrow$ Shock input},\\
\sin(\omega t) & \textrm{$\Rightarrow$ Acceleration wave input},
\end{cases}
\en
where the angular frequency $\omega (>0)$ is a  constant, and $\Theta(\zeta)$ denotes the Heaviside unit step function.  This hIBVP, we observe, is  known as a (acoustic) signaling problem~\cite[p.~189]{C98}.  


\subsection{Singular surface results}

As in Ref.~\cite[\S 4]{chen1973}, we define the amplitude of the jump  in a function $\mathfrak{F}=\mathfrak{F}(z,t)$ across a  singular surface $z=\Sigma(t)$  as
\begin{equation}
\lshad \mathfrak{F} \rshad :=\mathfrak{F}^{-}-\mathfrak{F}^{+},
\end{equation}
where $\mathfrak{F}^{\mp} := \lim_{z\to \Sigma(t)^{\mp}}\mathfrak{F}(z,t)$ are assumed to exist, and where a  \lq $+$' superscript corresponds to the region into which $\Sigma$ is advancing while a \lq $-$' superscript corresponds to the region behind $\Sigma$. Physically,  the surface $z=\Sigma(t)$ represents a wavefront.


Using Sys.~\eqref{Sys:atmos_psi} and the Rankine--Hugoniot conditions~(\cite[\S 6.3]{Bland88}, \cite{LeVeque02}),  in the shock case, and Maxwell's theorem~\cite[p.~494]{tt60}, in the acceleration wave case,  leads us to  the  ODE  
\be
\left( \frac{\rd \Sigma (t) }{\rd t}\right)^2 = V^{2}(t),
\en
 to be solved subject to the IC $\Sigma(0) =0$, with only the positive (i.e., `$+$' sign case) solution retained. Here,  $V(t)= c_{\rm a}(\Sigma(t))$ is the speed at which  $\Sigma(t)$  propagates (upward) along the $+z$-axis.

Applying the tools of singular surface theory~(see, e.g., Refs.~\cite[\S 6.9]{Bland88}, \cite{chen1973,S11,tt60}) to Sys.~\eqref{Sys:atmos_psi}, we are able to determine  the evolution of $\lshad w \rshad$, in the shock case, and $\lshad w_{z} \rshad$ and $\lshad w_{t} \rshad$, in the  acceleration wave case, for all fluids described by this system, where, as per hIBVP~\eqref{IBVP:Atmos_Gen},  $w^{+}  = 0$ is hereafter assumed.

For the shock input signal case, $\lshad w \rshad_{0} \neq  0$ and we find  that  
\be\label{eq:Shock_Amp_Gen_I}
\lshad w \rshad = \lshad w \rshad_{0}\sqrt{\frac{V(0)}{V(t)}}\exp\!\left\{-\,\tfrac{1}{2}\int_{0}^{t}\left[\frac{\varrho_{\rm a}^{\prime}(\Sigma(\zeta))}{\varrho_{\rm a}(\Sigma(\zeta))}\right]V(\zeta)\, \rd \zeta \right\}\!,
\en
which on carrying out the integration and simplifying becomes
\be\label{eq:Shock_Amp_Gen}
\lshad w \rshad = \lshad w \rshad_{0}\sqrt{\frac{V(0)}{V(t)}}\sqrt{\frac{\varrho_{\rm a}(\Sigma(0))}{\varrho_{\rm a}(\Sigma(t))}}.
\en
In Eq.~\eqref{eq:Shock_Amp_Gen}, $\lshad w \rshad_{0}$ denotes the value of $\lshad w \rshad$ at time $t=0$, which, in the case of  hIBVP~\eqref{IBVP:Atmos_Gen}, has the value $\lshad w \rshad_{0}= W_{0}$, and we note that $V(0)=c_{\rm a}(\Sigma(0))$. (While the notation has been suppressed, the reader should keep in mind that $\lshad w \rshad$, and all other jump relations derived hereafter, are explicitly functions of $t$.)

For the acceleration wave case,  $\lshad w \rshad_{0}=0$, but  $\lshad w_{t} \rshad_{0} \neq 0$; nevertheless, we get a similar amplitude expression:
\be\label{eq:Accel_Amp_Gen_I}
\lshad w_z \rshad = -\lshad w_t \rshad_{0}\frac{\sqrt{V(0)}}{V^{3/2}(t)}\exp\!\left\{-\,\tfrac{1}{2}\int_{0}^{t}\left[\frac{\varrho_{\rm a}^{\prime}(\Sigma(\zeta))}{\varrho_{\rm a}(\Sigma(\zeta))}\right]V(\zeta)\, \rd \zeta \right\}\!,
\en
which we can immediately reduce to
\be\label{eq:Accel_Amp_Gen}
\lshad w_z \rshad = -\lshad w_t \rshad_{0}\frac{\sqrt{V(0)}}{V^{3/2}(t)} \sqrt{\frac{\varrho_{\rm a}(\Sigma(0))}{\varrho_{\rm a}(\Sigma(t))}}.
\en
Here, $\lshad w_t \rshad_{0}$ denotes the value of $\lshad w_t \rshad$ at time $t=0$; in the case of  hIBVP~\eqref{IBVP:Atmos_Gen}, it has the  value  $\lshad w_t \rshad_{0}=\omega W_{0}$.  Also, in obtaining Eq.~\eqref{eq:Accel_Amp_Gen_I}  we have once again made use of Maxwell's theorem, this time in the form $\lshad w_t \rshad = - V(t)\lshad w_z \rshad$.

Lastly, it should be noted that the expressions derived in this subsection apply \emph{only} for times prior to the time the wavefront in question encounters a boundary, assuming one is present.

\subsection{Exponential density profile}\label{sect:Exp}

We begin with the simplest case of $\varrho_{\rm a}(z)$, i.e.,  of the so-called `exponential atmosphere':
\be\label{eq:Exp_density}
\varrho_{\rm a}(z) = \varrho_{0}\exp(-z/H) \qquad (z>0).
\en
Here, $H = c_{0}^{2}/(\gamma g)$ is the height of the `homogeneous atmosphere' under the assumption that $\vartheta_{\rm a}(z)$ is constant, specifically, that $\vartheta_{\rm a}(z)=\vartheta_{0}$ for all $z>0$~\cite[p.~542]{Lamb45}. 

In the case of Eq.~\eqref{eq:Exp_density}, then, Eq.~\eqref{eq:statical_rel} implies that $p_{\rm a}(z) = p_{0}\exp(-z/H)$, and Eq.~\eqref{eq:EoM_atmos_gas} and~\eqref{IBVP:Atmos_Gen_EoM} reduce to
\be\label{eq:EoM_atmos_gas_Exp}
w_{tt} =  c_{0}^{2}w_{zz} -\gamma g w_{z}.
\en

On replacing Eq.~\eqref{IBVP:Atmos_Gen_EoM} and $\lambda$ with Eq.~\eqref{eq:EoM_atmos_gas_Exp} and $\infty$, respectively, hIBVP~\eqref{IBVP:Atmos_Gen} becomes
\begin{subequations}\label{IBVP:Atmos_Exp}
\begin{align}
\label{IBVP:Atmos_Exp_EoM}
& w_{tt} =  c_{0}^{2}w_{zz} -\gamma g w_{z}, \quad (z,t)\in (0, \infty)\times (0, \infty),\\
& w(0,t) = W_{0}\Theta (t)f(t),\quad |\overline{w}(\infty, s)| < \infty, \quad  t, s >0,\\
& w(z,0) = 0, \quad  w_{t}(z,0) = 0, \quad  z \in (0, \infty).
\end{align}
\end{subequations}
Exact solutions to hIBVP~\eqref{IBVP:Atmos_Exp}  for both cases of $f(t)$  can be determined using the Laplace transform; see Ref.~\cite[\S 82]{CJ63}, wherein the exact solution for the case $f(t)=1$ is given.

In the case of hIBVP~\eqref{IBVP:Atmos_Exp},  $\Sigma (t) = c_{0}t$, meaning that $V(t) = c_{0}$; thus,  from Eqs.~\eqref{eq:Shock_Amp_Gen} and~\eqref{eq:Accel_Amp_Gen}, one finds that  the resulting shock and acceleration wave amplitudes are given by
\begin{align}
\lshad w \rshad &= W_{0}\exp[c_{0}t/(2H)] \qquad (t > 0), \label{eq:Sock_amp_Atmos_Exp}\\
\lshad w_{z} \rshad &= - \omega c_{0}^{-1}W_{0}\exp[c_{0}t/(2H)] \qquad (t > 0), \label{eq:Accel_amp_Atmos_Exp}
\end{align}
respectively, with both jumps occurring across $\Sigma(t)=c_{0}t$.  Note that, like the plane wave solution to Eq.~\eqref{eq:EoM_atmos_gas_Exp} derived by Whitham~\cite[p.~160]{W74} (see also Lamb~\cite[p.~543]{Lamb45}), the solutions to which these these jump expressions correspond are only valid for $z \ll H$ since their magnitudes also increase without bound as $t\to \infty$. [This is easily established in the case of Eq.~\eqref{eq:Sock_amp_Atmos_Exp} because $w^{-}(t)=\lshad w \rshad$.]

\begin{remark}\label{rmk:Nonlinear} 
It is apt to mention the \emph{nonlinear}  acceleration wave analysis performed by Walsh~\cite{Walsh69}, who also considered vertically-propagating acoustic wavefronts in the atmosphere; Walsh, however, took   $\varrho_{\rm a}(z)$ \emph{and} $\vartheta_{\rm a}(z)$ as exponential functions.
\end{remark}

\subsection{Algebraic density profile}

We now consider a variant of the  profile in~\cite[\S 310]{Lamb45}, viz.,
\be\label{eq:Algebraic_density}
\varrho_{\rm a}(z) = \varrho_{0}(1-z/\ell)^{\chi} \qquad (0 < z <\ell),
\en
which, with the aid of Eq.~\eqref{eq:statical_rel}, yields
\be\label{eq:Algebraic_pressure}
p_{\rm a}(z) = p_{0}(1-z/\ell)^{\chi+1}.
\en
Here,  we must have $ \ell  := (\chi+1) H$, where $\chi(\geq 0)$ is a (known) constant, to ensure that $\vartheta_{\rm a}^{\prime}(z)$  is always \emph{negative} and constant-valued~\cite[p.~545]{Lamb45}, and $p_{0}   = g\varrho_{0} H$.  Thus, for this class of density profiles Eq.~\eqref{eq:EoM_atmos_gas}  becomes
\be\label{eq:EoM_atmos_gas_Algebraic}
w_{tt} =  c_{\rm a}^{2}(z)w_{zz} -c_{0}^{2}\ell^{-1}(\chi+1) w_{z} \qquad [c_{\rm a}^{2}(z) = c_{0}^{2}(1-z/\ell)].
\en

Now replacing Eq.~\eqref{IBVP:Atmos_Gen_EoM} with Eq.~\eqref{eq:EoM_atmos_gas_Algebraic} and setting $\lambda = \ell$, IBVP~\eqref{IBVP:Atmos_Gen} becomes
\begin{subequations}\label{IBVP:Atmos_Algebraic1}
\begin{align}
\label{eq:IBVP:Atmos_Algebraic1}
& w_{tt}= c_{0}^{2}(1-z/\ell)w_{zz} - c_{0}^{2}\ell^{-1}(\chi+1) w_{z}, \quad (z,t)\in (0, \ell)\times (0, \infty);\\
& w(0,t) = W_{0}\Theta (t)f(t),\quad |\overline{ w}(\ell,s)| < \infty, \quad  t, s>0;\\
& w(z,0) = 0, \quad  w_{t}(z,0) = 0, \quad  z \in(0, \ell).
\end{align}
\end{subequations}
Introducing  the following dimensionless variables:
\be\label{eq:ND_qts}
W=w/W_{0}, \qquad Z=z/\ell, \qquad T= t/(\ell/c_{0}),
\en
 hIBVP~\eqref{IBVP:Atmos_Algebraic1} simplifies to 
\begin{subequations}\label{IBVP:Atmos_Algebraic2}
\begin{align}
\label{eq:IBVP:Atmos_Algebraic2}
& W_{TT}= (1-Z)W_{ZZ} - (\chi+1) W_{Z},  \quad (Z,T)\in (0, 1)\times (0, \infty);\\
& W(0,T) = \Theta(T)f(T),\quad |\overline{W}(1,s)| < \infty, \quad  T, s > 0;\\
& W(Z,0) = 0, \quad  W_{T}(Z,0) = 0, \quad  Z \in (0, 1).
\end{align}
\end{subequations}
Here, on setting $\Omega := \omega \ell/c_{0}$, we now have
\be\label{eq:f_ND}
f(T)=\begin{cases}
1 & \textrm{$\Rightarrow$ Shock input},\\
\sin(\Omega T) & \textrm{$\Rightarrow$ Acceleration wave input}.
\end{cases}
\en

Applying the Laplace transform to Eq.~\eqref{eq:IBVP:Atmos_Algebraic2} and the left-BC, and then using the ICs,  hIBVP~\eqref{IBVP:Atmos_Algebraic2} is reduced to the following boundary value problem (BVP) in the  transform domain:
\begin{subequations}\label{BVP:Atmos_Algebraic2}
\begin{align}
& (1-Z)\overline{W}_{ZZ} - (\chi+1)\overline{W}_{Z} = s^2\overline{W}, \quad Z\in (0, 1),\\
& \overline{W}(0,s) = \overline{f}(s),\quad |\overline{W}(1,s)| < \infty, \quad s>0.
\end{align}
\end{subequations}
This subsidiary equation can be transformed into a Bessel-type ODE (see Ref.~\cite[p.~546]{Lamb45}), after which it is not difficult to obtain the exact (transform-domain) solution
\be\label{eq:Lap-sol_Atmos_Algebraic}
\overline{W}(Z,s)=\overline{f}(s)\!\left[\frac{I_{\chi}(2s\sqrt{1-Z}\,)}{I_{\chi}(2s)(1-Z)^{\chi/2}} \right]\!,
\en
where it should be noted that
\be
\lim_{Z \to 1}\overline{W}(Z,s)= \frac{\overline{f}(s)}{\Gamma (\chi+1)}\left[\frac{s^{\chi} }{I_{\chi}(2s)}\right]\!.
\en
Here, $I_{\varsigma}(\zeta)$ and $\Gamma (\zeta)$ denote the modified Bessel function of the first kind of order $\varsigma$ and the gamma function, respectively.  As page limitations prevent us from doing so,  we leave the inversion of Eq.~\eqref{eq:Lap-sol_Atmos_Algebraic}, for both cases of $\overline{f}(s)$, to the reader.

Now, using Ref.~\cite[Eq.~(9.7.1)]{Abr65}, it can be shown that Eq.~\eqref{eq:Lap-sol_Atmos_Algebraic} admits the asymptotic expansion
\begin{multline}\label{eq:Lap-sol_Atmos_Algebraic_Asy}
\overline{W}(Z,s) \sim \frac{\overline{f}(s) \exp[-2s(1-\sqrt{1-Z}\,)]}{\sqrt{(1-Z)^{\chi+1/2}}}\\
\times \left[1+\frac{4\chi^2-1}{16 s}\left(1- \frac{1}{\sqrt{1-Z}} \right) + \cdots \right]
\quad (s \to \infty).
\end{multline}
In the remainder of this section, we shall limit our focus to $0 < T < T_{\rm R}$, where $T_{\rm R}$ is the time that the wavefront created by the signal in question \emph{first} reaches the right boundary (i.e., $Z=1$).

For the shock   and acceleration wave  cases, then, inverting Eq.~\eqref{eq:Lap-sol_Atmos_Algebraic_Asy} term-by-term yields the small-$T$ approximations:
\begin{multline}
W(Z,T) \approx \frac{\Theta[T-2(1-\sqrt{1-Z}\,)]}{\sqrt{(1-Z)^{\chi+1/2}}} \\
\times \left\{1+ \frac{(4\chi^2-1)[T-2(1-\sqrt{1-Z}\,)]}{16}\left(1- \frac{1}{\sqrt{1-Z}} \right)  \right\}
\\ (T \ll T_{\rm R}),
\end{multline}
for which $\overline{f}(s) = 1/s$ was used, and
\begin{multline}
W(Z,T) \approx \frac{\Omega \Theta[T-2(1-\sqrt{1-Z}\,)]}{\sqrt{(1-Z)^{\chi+1/2}}} \\
\times \Bigg\{T-2(1-\sqrt{1-Z}\,)+ \frac{(4\chi^2-1)[T-2(1-\sqrt{1-Z}\,)]^2}{32}\\
\times \left(1- \frac{1}{\sqrt{1-Z}} \right)  \Bigg\} \quad (T \ll T_{\rm R}),
\end{multline}
for which $\overline{f}(s) = \Omega/(s^2+\Omega^2) \sim  \Omega s^{-2}(1-\Omega^2/s^2 +\cdots)$ was used, respectively.  

If we now replace $Z$ with $\sigma(T)$ in  the  argument of the Heaviside function in these approximations, set the  argument expression to  zero, and then solve for $\sigma(T)$, we find that
\be\label{eq:algebraic_early_sigma}
\sigma (T) = T-\tfrac{1}{4}T^2 \qquad (0 < T < 2),
\en
from which it follows that $V(T) = 1-T/2$ and $T_{\rm R}=2$; here, $\sigma$ denotes the dimensionless version of $\Sigma$.  Thus, using the general expressions given in Eqs.~\eqref{eq:Shock_Amp_Gen} and~\eqref{eq:Accel_Amp_Gen}, the shock and acceleration wave amplitudes stemming from the density profile given in Eq.~\eqref{eq:Algebraic_density} are found  to be
\begin{align}
\lshad W \rshad &= \frac{1}{(1-T/2)^{\chi+1/2}} \qquad (0 < T < 2),\label{eq:Sock_amp_Atmos_Algebraic}\\
\lshad W_{Z} \rshad &= \frac{-\Omega}{(1-T/2)^{\chi+3/2}}  \qquad (0 < T < 2),\label{eq:Accel_amp_Atmos_Algebraic}
\end{align}
respectively, with both jumps occurring across $Z=\sigma (T)$.  In the shock case we once again see amplitude blow-up, but now as $T \to T_{\rm R}(=2)$.  In contrast, the acceleration wave case is more interesting.  This is because, like the nonlinear version of this case of  hIBVP~\eqref{IBVP:Atmos_Algebraic2} involving a homogeneous gas (see, e.g., Ref.~\cite{QJMAM}), Eq.~\eqref{eq:Accel_amp_Atmos_Algebraic} exhibits wavefront steepening (i.e., `shocking-up') as $T \to 2$; however, unlike that of the former, the solution profile corresponding to Eq.~\eqref{eq:Accel_amp_Atmos_Algebraic} \emph{also} blows-up as $T \to 2$.  These behaviors are illustrated below in Fig.~\ref{fig:algebraic}, wherein snapshots in the evolution of $\lshad W \rshad$  and $\lshad W_{Z} \rshad$ are presented.

\begin{remark}~\label{Rem_special-cases_Algebraic}
Of particular interest are the special cases $\chi = 0$ and  $\chi =(\gamma -1)^{-1}$. For these values of $\chi$,  Eq.~\eqref{eq:Algebraic_density} becomes
\be
\varrho_{\rm a}(z) = \varrho_{0}
\begin{cases} 
1, & \chi = 0,\\
(1-z/H_{\rm h})^{\frac{1}{\gamma -1}}, & \chi = (\gamma -1)^{-1},
\end{cases}
\en
where  $H_{\rm h} := \gamma H/(\gamma -1)$.  Here,   $\chi = 0, (\gamma -1)^{-1}$ correspond to $\ell= H, H_{\rm h}$, respectively, where    $H_{\rm h} > H$,  and the subscript `h' signifies that in this case  the  ambient state is \emph{homentropic}\footnote{Meaning that $\eta_{\rm a}(z) =\eta_{0}$, for all $z\in (0, H_{\rm h})$, which follows from Eq.~\eqref{eq:entropy_gas} and  the fact that   Eq.~\eqref{eq:Algebraic_pressure} can be written as $p_{\rm a}(z) = p_{0}[\varrho_{\rm a}(z)/\varrho_{0}]^{\gamma}$ when $\chi=(\gamma -1)^{-1}$. This case defines an atmosphere in `convective equilibrium'~\cite[p.~546]{Lamb45}.}.   Also,  $\chi=0$ implies that  
$\vartheta_{\rm a}^{\prime}(z)= -\beta := -g/(c_p-c_v)$, for all $z\in (0, H)$~\cite[p.~545]{Lamb45}; similarly, $\chi=(\gamma -1)^{-1}$ implies that  $\vartheta_{\rm a}^{\prime}(z)= -\beta_{1} := -g/c_{p}$, for all $z\in (0, H_{\rm h})$~\cite[p.~546]{Lamb45}.
\end{remark}

\begin{remark}\label{rmk:boley} The shock and acceleration wave results presented in this subsection can also be obtained by applying the theorem given in Ref.~\cite[\S 4]{BH68} to Eq.~\eqref{eq:Lap-sol_Atmos_Algebraic_Asy} and then, in the case of  the latter, employing Maxwell's theorem.  
\end{remark}

\subsection{Numerical results: algebraic density profile}

In this subsection we compute and plot the velocity field solution, for several cases of the algebraic density profile, by numerically inverting the Laplace domain solution [Eq.~\eqref{eq:Lap-sol_Atmos_Algebraic}] using the formula 
\begin{multline}
W(Z,T) \approx \frac{\exp(4.7)}{T}\left\{\frac{1}{2}\overline{W}\left(Z,\frac{4.7}{T}\right) \right. \\ \left. +\Re\left[ \sum_{m=1}^{M} (-1)^m \overline{W}\left(Z,\frac{4.7+\ri m \pi}{T} \right) \sinc\left(\frac{m\pi}{M}\right)\right] \right\}\!,
\label{eq:Tzou}
\end{multline}
where $T>0$.  Equation~\eqref{eq:Tzou} is a modified version of Tzou's~\citep{Tzou97} Riemann-sum inversion approximation, obtained by introducing Lanczos' `$\sigma$-factors'~\cite{Lanczos}, i.e., the $\sinc(m\pi/M)$, where 
\be
\sinc (\xi) := 
\begin{cases}
\xi^{-1}\sin (\xi), & \xi \neq 0,\\
1, & \xi = 0,
\end{cases}
\en
into the latter. This is done to  reduce the Gibbs phenomenon  in Fourier series, such as Eq.~\eqref{eq:Tzou}, near discontinuities in the function being approximated,  as in, e.g., our shock solution [i.e., the case $\overline{f}(s)=1/s$],  \emph{without} affecting the series' convergence.  Here, we have set $M=1,500$; the number `4.7' and the quantity to which it is assigned are discussed in Ref.~\citep[p.~41]{Tzou97}.

\begin{figure}[h]
    \centering
    \includegraphics[width=0.475\textwidth]{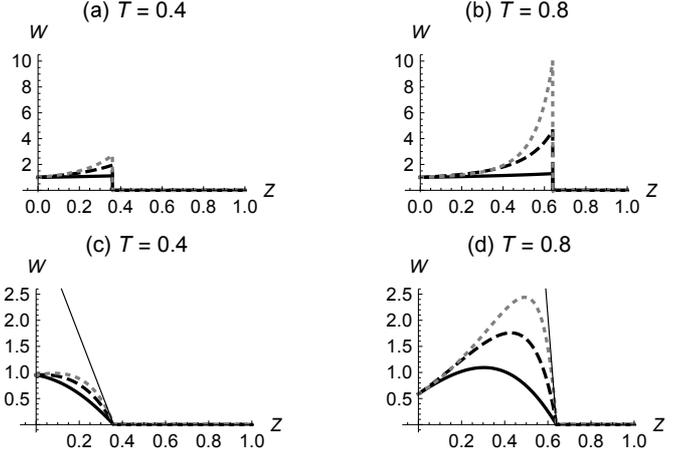}
    \caption{Evolution of a velocity shock wave [panels (a) and (b)] and a velocity acceleration wave [panels (c) and (d)] in an atmosphere with an algebraic density profile for $\chi=0$ (solid black curves), $\chi=2.5$ (dashed black curves) and $\chi=4$ (dashed gray curves). Note the different vertical scales in panels (a) and (b) versus (c) and (d). For the acceleration wave case, $\Omega=\pi$, and the thin black slanted lines in panels (c) and (d) are the  theoretical predictions, based on Eq.~\eqref{eq:Accel_amp_Atmos_Algebraic}, of the wavefront tangents for the case $\chi=4$.}
    \label{fig:algebraic}
\end{figure}

Figure~\ref{fig:algebraic} shows the evolution of a shock wave [panels (a) and (b)] and an acceleration wave [panels (c) and (d)] based on  Eq.~\eqref{eq:Lap-sol_Atmos_Algebraic}  for three choices of the algebraic exponent: $\chi = 0, 2.5, 4$, where  $\chi = 2.5$ corresponds to air, for which $\gamma = 1.4$, when the ambient state is homentropic  (see \textbf{Remark~\ref{Rem_special-cases_Algebraic}}). As suggested by Eq.~\eqref{eq:algebraic_early_sigma}, the location of the wavefront is independent of  $\chi$. The shock and acceleration wave amplitudes grow algebraically without bound, however, as indicated by Eqs.~\eqref{eq:Sock_amp_Atmos_Algebraic} and \eqref{eq:Accel_amp_Atmos_Algebraic}. In Fig.~\ref{fig:algebraic}, we have also included the  acceleration wave's wavefront location and amplitude, i.e., the straight lines plotted from $\lshad W_Z \rshad (T)[Z-\sigma(T)]$, where the acceleration wave amplitude  is given by Eq.~\eqref{eq:Sock_amp_Atmos_Algebraic} and the wavefront location   by Eq.~\eqref{eq:algebraic_early_sigma}, which show excellent agreement with the solution obtained via numerical inversion.

\section{Propagation in a fluid with a periodic density profile   in the absence of external body forces}\label{sect:Periodic}

\subsection{Linearized system and equation of motion}

In this section  we assume 1D planar propagation along the $x$-axis; i.e., we take ${\bf u}=(u(x,t),0,0)$ and  ${\bf b} = (0,0,0)$, and  observe that in this setting $p$ and $\varrho$ are, like $u$, both functions of  $x$ and $t$ only.  Again eliminating  $D\varrho/Dt$ between Eq.~\eqref{eq:cont_1}  and~\eqref{eq:EoS_1},    but now setting ${\mathsf{p}}(x,t) = p(x,t) -p_{\rm a}(x)$, it is  a straightforward matter to linearize this special case of Sys.~\eqref{sys:Comp} and express it as%
\begin{subequations} \label{sys:b0}
\begin{align}
	\varrho_{\rm a}(x)u_{t} + {\mathsf{p}}_{x} &= -p_{\rm a}^{\prime}(x),\\
	{\mathsf{p}}_{t}+ A_{\rm a}(x)u_{x} &= -p_{\rm a}^{\prime}(x)u,
\end{align}
\end{subequations}
where in this section a prime denotes $\rd /\rd x$.   Here, we observe that if one seeks to determine only $u$ and/or $p$, then the ${\bf b} = (0,0,0)$ special case of the 1D version of Sys.~\eqref{sys:Comp} can always be reduced to  a  \emph{two-equation} system.

Since  the ambient state values of the field variables must also satisfy this system, it follows that $p_{\rm a}(x)$ is  necessarily  a constant, specifically, $p_{\rm a}(x) = p_{\rm r}$, where $p_{\rm r}$ is a reference value of $p_{\rm a}(x)$; hence, $p_{\rm a}^{\prime}(x)=0$ and, on setting $\varphi(x,t) := {\mathsf{p}}(x,t)/\varrho_{\rm a}(x)$,  Sys.~\eqref{sys:b0} can be  recast as 
\begin{subequations} \label{Sys:b0_varphi}
\begin{align}
	u_{t} + \varphi_{x} &= -[\varrho_{\rm a}'(x)/ \varrho_{\rm a}(x)] \varphi, \label{eq:flux_phi}\\
	\varphi_{t}+ c_{\rm a}^{2}(x) u_{x} &= 0,\label{eq:conv_phi}
\end{align}
\end{subequations}
where  we note that  $c_{\rm a}^{2}(x)= A_{\rm a}(x)/\varrho_{\rm a}(x)$ in the present section.


On eliminating  $\varphi$  between the equations of Sys.~\eqref{Sys:b0_varphi}, the corresponding equation of motion  is easily shown to be 
\be\label{eq:WaveEq_1}
u_{tt}-c_{\rm a}^{2}(x)u_{xx} =0.
\en
If we now assume the periodic density profile
\be\label{eq:periodic_rho}
\varrho_{\rm a}(x) = \varrho_{\rm r}[1+\epsilon \cos(k \pi x/L)] \qquad (0<x<L), 
\en
and that $A_{\rm a}(x)$ is  constant\footnote{In the case of   perfect gases this need \emph{not} be assumed since
$A_{\rm a}(x)=\gamma p_{\rm r}$ holds exactly; i.e., $A_{\rm r}=\gamma p_{\rm r}$, and  thus $c_{\rm r}^{2}=\gamma p_{\rm r}/\varrho_{\rm r}$,  for perfect gases.}, i.e., $A_{\rm a}(x)=A_{\rm r}$, then  it follows that
\be\label{eq:c_1}
c_{\rm a}^{2}(x) = c_{\rm r}^2[1+\epsilon \cos(k\pi x/L)]^{-1} \qquad (k\in \mathbb{N}).
\en
Here,  $\varrho_{\rm r}$ and $A_{\rm r}$ represent reference values of $\varrho_{\rm a}(x)$ and $A_{\rm a}(x)$, respectively;   $\epsilon \in (0,1)$ is  a (dimensionless) parameter;  $c_{\rm r}^{2}=A_{\rm r}/\varrho_{\rm r}$; and  $L$ is  a characteristic length of the domain.

\subsection{Formulation of IBVP and singular surface results}\label{sec:periodic_SST}

We now consider the following IBVP involving Eq.~\eqref{eq:WaveEq_1} with $c_{\rm a}$ given by \eqref{eq:c_1}, and wherein $U_0(> 0)$ is a (known) constant:
\begin{subequations}
\begin{align}
& u_{tt}-c_{\rm a}^{2}(x)u_{xx} =0, \quad (x,t)\in (0, L)\times (0,  t_{\rm f}),\\
& u(0,t) = U_{0}\Theta (t)f(t),\quad u_{x}(L,t) = 0, \quad  t\in (0, t_{\rm f}) ,\\
& u(x,0) = 0, \quad  u_{t}(x,0) = 0, \quad  x \in(0, L).
\end{align}\label{eq:WE_IBVP_Signal1}
\end{subequations}
As the shock and acceleration wave results presented below are only valid for such times, we  have limited our focus to $0 < t < t_{\rm f}$, where $t_{\rm f}$ is the time at which the wavefront of the input signal in question  \emph{first} reaches the right boundary (i.e., $x=L$). 

Notwithstanding the fact that it involves a linear PDE and linear BCs, at present,  there appears  little hope of obtaining  an analytical solution to this IBVP.  Accordingly, we must turn to numerical methods if further progress is to be achieved.   

To this end, we introduce  the following dimensionless variables:
$U=u/U_{0}$, $X=x/L$, and $T= t/(L/c_{\rm r})$.  With these substitutions, our IBVP is reduced to 
\begin{subequations}\label{eq:WE_IBVP_Signal2}
\begin{align}
& U_{TT}= \mathcal{C}_{\rm a}^{2}(X)U_{XX}, \quad (X,T)\in (0, 1)\times (0, T_{\rm f}) \label{EoM:Periodic_Signal2},\\
& U(0,T) = \Theta(T)f(T), \quad U_{X}(1,T) = 0, \quad  T\in (0, T_{\rm f}),\\
& U(X,0) = 0, \quad  U_{T}(X,0) = 0, \quad  X \in (0, 1).
\end{align}
\end{subequations}
Here,  $\mathcal{C}_{\rm a}(X) = c_{\rm a}(LX)/c_{\rm r}= [1+\epsilon \cos(k \pi X)]^{-1/2}$ and  $f(T)$ is again given by Eq.~\eqref{eq:f_ND}, but in this section $\Omega$ is defined as $\Omega : = \omega L/c_{\rm r}$.  Also,  $T_{\rm f}$, the dimensionless version of $t_{\rm f} $, is given by
\be\label{eq:Tf}
T_{\rm f} = \frac{2\sqrt{1+\epsilon}}{k \pi} \left[E\left( \tfrac{1}{2}k\pi\, \Bigg{|}\, \frac{2\epsilon}{1+\epsilon} \right) \right] =    \frac{2\sqrt{1+\epsilon}}{\pi} \left[E\left( \frac{2\epsilon}{1+\epsilon} \right) \right]\!,
\en 
where $E(\zeta\,|\, \mathsf{m})$ and  $E(\mathsf{m})$ are the incomplete and complete, respectively, elliptic integrals of the second kind with parameter $\mathsf{m}\in (0,1)$~\cite{Abr65,Boyd12}; i.e.,  $\Upsilon (T_{\rm f}) =1$, where $\Upsilon$  is implicitly given by
\be\label{eq:sigma_exact}
T= \frac{2\sqrt{1+\epsilon}}{k \pi} \left[E\left( \tfrac{1}{2}k\pi \Upsilon\, \Bigg{|}\, \frac{2\epsilon}{1+\epsilon} \right) \right]\!,
\en 
and where $X=\Upsilon (T)$ is the location of the wavefront during its \emph{initial} transit of the interval 
$0<X<1$.

Equation~\eqref{eq:sigma_exact}, we observe, was obtained by integrating, subject to the IC $\Upsilon(0) =0$, the `$+$' sign case of  the ODE  
\be\label{eq:shock_front_ODE}
\left( \frac{\rd \Upsilon }{\rd T}\right)^2 = \mathcal{C}_{\rm a}^{2}(\Upsilon (T)) = \frac{1}{1+\epsilon \cos(k\pi \Upsilon(T))},
\en
where  the (dimensionless)  speed at which  $\Upsilon (T)$ propagates (to the right) along the $+X$-axis  is  $\mathcal{U}(T)= \mathcal{C}_{\rm a}(\Upsilon(T))$.

In the case of IBVP~\eqref{eq:WE_IBVP_Signal2}, then, the  shock and acceleration wave amplitudes are, using Eqs.~\eqref{eq:Shock_Amp_Gen} and~\eqref{eq:Accel_Amp_Gen}, found to be
\begin{align}
\lshad U \rshad &= \frac{(1+\epsilon)^{1/4}}{[1+\epsilon\cos(k\pi\Upsilon(T))]^{1/4}}, \quad  T\in (0, T_{\rm f}),\label{eq:Shock_Amp_b0}\\[2mm]
\lshad U_X \rshad &= -\Omega \Big\{(1+\epsilon) [1+\epsilon\cos(k\pi\Upsilon(T))]\Big\}^{1/4}, \,\,  T\in (0, T_{\rm f}),\label{eq:Accel_Amp_b0}
\end{align}
respectively, with both jumps occurring across $X =\Upsilon (T)$. 

\begin{remark}\label{rmk:epsilon}  To handle the case $\epsilon \in (-1,0)$, one must modify Eqs.~\eqref{eq:Tf} and~\eqref{eq:sigma_exact}  in accordance with  Ref.~\cite[Eq.~(17.4.18)]{Abr65}.
\end{remark}

\subsection{Approximations relating to IBVP~\eqref{eq:WE_IBVP_Signal2}}

On expanding Eq.~\eqref{eq:sigma_exact} for small-$\Upsilon$ we find that 
\be\label{eq:sigma_expand-sigma}
\Upsilon -\frac{1}{12}\left(\frac{\epsilon\pi^2k^2}{1+\epsilon} \right)\Upsilon^3 +\mathcal{O}(\Upsilon^5) = (1+\epsilon)^{-1/2}T \quad (0< \Upsilon < 1).
\en 
If we now neglect  terms of $\mathcal{O}(\Upsilon^5)$ and apply $\rd/\rd T$ to both sides,  then Eq.~\eqref{eq:sigma_expand-sigma} can be solved for $\mathcal{U}(T)$ to yield
\be\label{eq:U_cubic-sigma}
\mathcal{U}(T) \approx \mathsf{U}(\epsilon)[1-(\Upsilon/\Upsilon^{*})^2]^{-1} \qquad [0< \Upsilon \ll \min(1, \Upsilon^{*})].
\en
Here,  $\mathsf{U}(\epsilon)= (1+\epsilon)^{-1/2}$ is the speed at which  both the acceleration and shock wavefronts  propagate in the \emph{homogeneous} (i.e., $k \to 0$) fluid case and  $\Upsilon=\Upsilon^{*}$, where  $\Upsilon^{*} := 2(k\pi)^{-1}\sqrt{(1+\epsilon)/\epsilon}$, is the value at which the (two) positive roots of the $\mathcal{O}(\Upsilon^3)$-based  (i.e., cubic polynomial) approximation coalesce into a single root of multiplicity two.  It must be stressed, however, that this `backwards in time bifurcation' is an artifact of the $\mathcal{O}(\Upsilon^3)$-based  approximation---one that  is \emph{not} exhibited by  Eq.~\eqref{eq:sigma_exact}.

Equation~\eqref{eq:sigma_expand-sigma} also makes clear that, to lowest order,
\be\label{eqs:apporx_Up}
\Upsilon (T) \sim \mathsf{U}(\epsilon) T \qquad (T \to 0),
\en
while from  Eq.~\eqref{eq:U_cubic-sigma} we find that $\mathcal{U}(T) > \mathsf{U}(\epsilon)$, for $T \ll \min(T_{\rm f}, T^{*})$,
 where $T^{*} := \tfrac{4}{3}(k\pi)^{-1}(1+\epsilon)/\sqrt{\epsilon}$ is the value of $T$ corresponding to $\Upsilon^{*}$. 

We leave it to the reader to solve the aforementioned cubic; see  Ref.~\cite[p.~17]{Abr65}.  However, he/she should be aware that it is only the (positive) root which tends to zero, as $T \to 0$, that approximates $\Upsilon (T)$, and then only for $T \ll \min(T_{\rm f}, T^{*})$.




Returning  to Eq.~\eqref{eq:sigma_exact}, we now expand it for small-$\epsilon$.  After neglecting terms of $\mathcal{O}(\epsilon^2)$ and simplifying, we obtain
\be\label{eq:Kepler2}
k\pi\Upsilon + \tfrac{1}{2}\epsilon \sin(k\pi \Upsilon) \approx k\pi T \qquad (\epsilon \ll 1),
\en
which we observe is an approximate version of \emph{Kepler's equation}~\cite{EWW} with  \emph{negative} eccentricity $e = -\epsilon/2$.  Using Ref.~\cite[Eq.~(6)]{EWW}, Eq.~\eqref{eq:Kepler2} may be solved to give
the  approximation
\be
\Upsilon(T) \approx T+\frac{2}{k\pi}\sum_{n=1}^{\infty}\frac{(-1)^{n}}{n} J_{n}(n\epsilon/2)\sin(n k\pi T) \quad (\epsilon \ll 1),
\en 
where  $J_{\varsigma}(\zeta)$ is the Bessel function of the first kind of order $\varsigma$.

As is readily established,  $T_{\rm f}\in (0, 1)$; and from this it follows that  $\mathcal{U}_{\rm avg}(T) > 1 >  \mathsf{U}(\epsilon)$, where $\mathcal{U}_{\rm avg}(T) =1/T_{\rm f}$. Moreover, as expanding the last expression in  Eq.~\eqref{eq:Tf} for small-$\epsilon$ reveals,
\be
T_{\rm f} \approx 1-\tfrac{\epsilon^2}{16}-\tfrac{15\epsilon^4}{1024}, \quad \mathcal{U}_{\rm avg}(T) \approx 1+\tfrac{\epsilon^2}{16}+\tfrac{19\epsilon^4}{1024}  \quad (\epsilon \ll 1),
\en
where terms of $\mathcal{O}(\epsilon^6)$ have been neglected.

\subsection{Numerical results: periodic density profile}

Our numerical approach to acoustic propagation in a periodic fluid medium in the absence of external body forces is based on the dimensionless version of the \emph{non-conservative} system \eqref{sys:b0}:%
\begin{subequations}\label{Sys:NC_dimles}
\begin{align}
	[\mathcal{C}_{\rm a}^2(X)]^{-1}U_{T} + P_{X} &= 0,\\
	P_{T} + U_{X} &= 0,
\end{align}
\end{subequations}
where $P = \mathsf{p}/(\varrho_\mathrm{r}c_\mathrm{r}U_0)$. 
To solve Sys.~\eqref{Sys:NC_dimles} numerically, subject to the stated ICs and BCs, we employ the modern extensible software package PyClaw~\cite{PyClaw1,PyClaw2}.

PyClaw is based on LeVeque's CLAWPACK~\cite{LeVeque02}. We employed the PyClaw solver based on the second-order-accurate wave propagation algorithm~\cite{L97}; see also Refs.~\cite{LeVeque99, LeVeque02}. The wave propagation method is a high-resolution shock-capturing scheme capable of handling non-conservative hyperbolic systems of PDEs. The PyClaw package and its Riemann solvers (for handling discontinuous solutions) have been benchmarked against other methods and exact solutions; thus, the numerical solutions shown below are robust, reproducible, and highly accurate. Specifically, we have employed a Python implementation of the variable coefficient acoustics Riemann solver~\cite{LeVeque99}. High-resolution shock-capturing schemes require limiters to resolve discontinuities; we employed the so-called monotonized central (MC) limiter~\cite{LeVeque02}. The simulations discussed below were performed using a computation grid of $20,000$ (for $k=5$ and $8$) and $60,000$ (for $k=15$ and $25$) cells on the domain $X \in[0,1]$. A second-order, adaptive time-stepping scheme was used, which maintained a `target' Courant--Friedrichs--Lewy (CFL) number of 0.9. At $X=0$, an inlet velocity BC was applied, while at $X=1$ a transmissive (i.e., extrapolation) BC was employed. These BCs were implemented using two ghost cells on each side of the computational domain (see Ref.~\cite{LeVeque02}) to maintain the overall second-order accuracy of the scheme.

\begin{figure}[h]  
	\centering
	\includegraphics[width=0.475\textwidth]{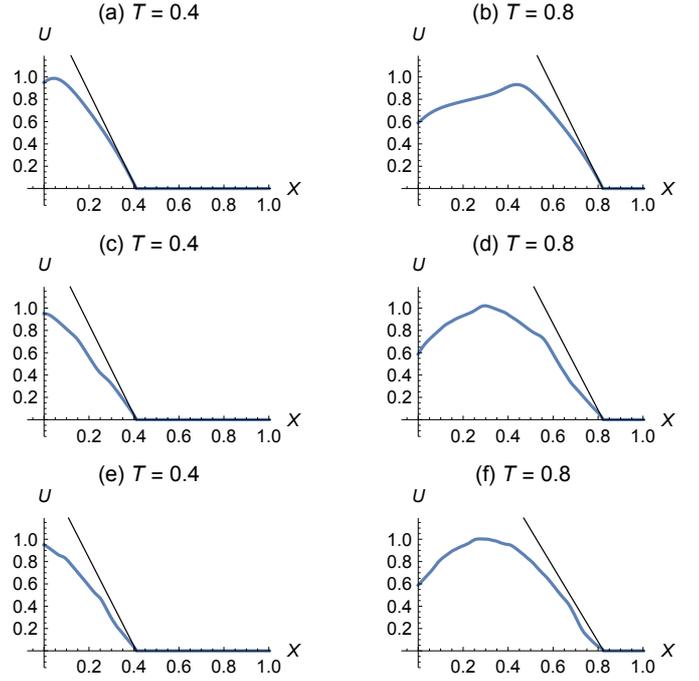}
	\caption{Evolution of an acceleration wave in $U(X,T)$ for  $\epsilon=0.7$, $\Omega=\pi$,  $k=5$ [panels (a), (b)], $k=15$ [panels (c), (d)], and $k=25$ [panels (e), (f)]. The thin black slanted lines are the theoretical prediction for the wavefront dynamics plotted from $\lshad U_X \rshad (T)[X-\Upsilon(T)]$, where  $\lshad U_X \rshad$ is given by Eq.~\eqref{eq:Accel_Amp_b0}.}
	\label{fig:aw_num}
\end{figure}

\begin{figure}[h!]
	\centering
	\includegraphics[width=0.475\textwidth]{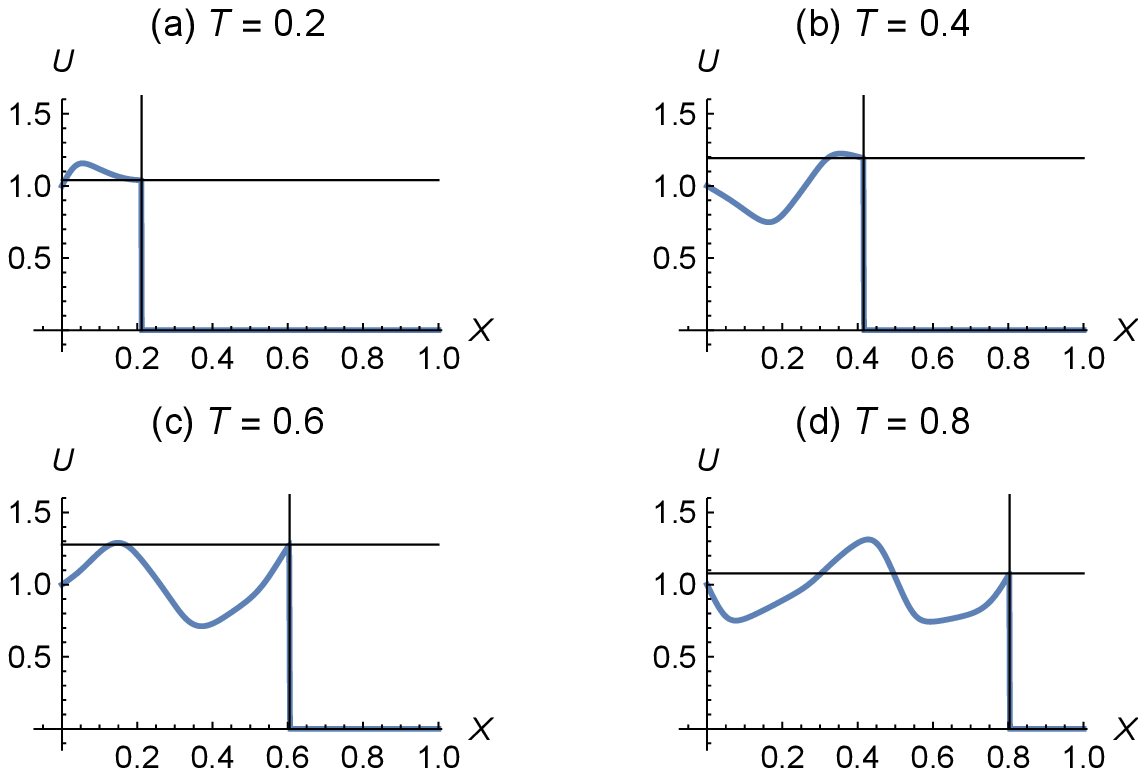}
	\caption{Evolution of a shock wave in $U(X,T)$ for $\epsilon=0.5$ and $k=8$. The thin horizontal and vertical lines in each panel are theoretical predictions for the shock amplitude $\lshad U \rshad $ and wavefront location $\Upsilon$ are given by Eq.~\eqref{eq:Shock_Amp_b0} and by the numerical integration of Eq.~\eqref{eq:shock_front_ODE}, respectively.}
	\label{fig:shock_num}
\end{figure}

\begin{figure}[h!]  
	\centering
	\includegraphics[width=0.475\textwidth]{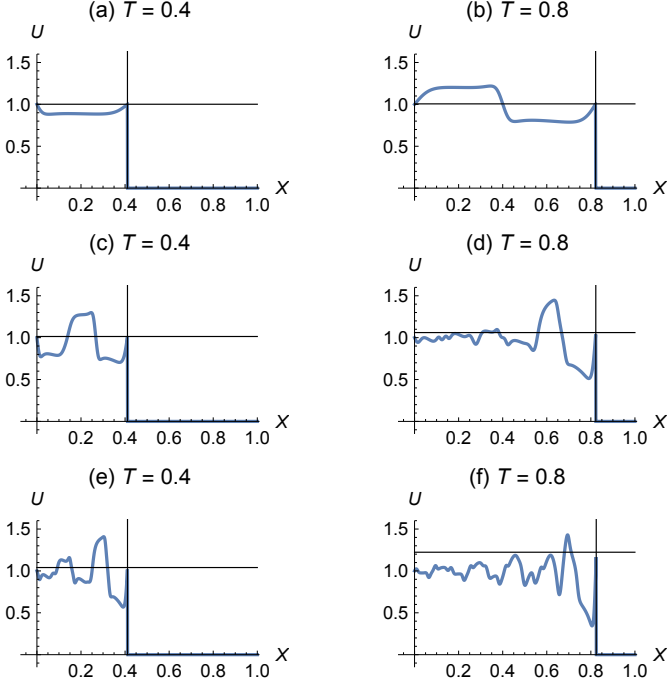}
	\caption{Evolution of a shock wave in $U(X,T)$ for $\epsilon=0.7$, $k=5$ [panels (a), (b)], $k=15$ [panels (c), (d)], and $k=25$ [panels (e), (f)]. The thin horizontal and vertical lines in each panel are theoretical predictions for the shock amplitude $\lshad U \rshad $ and wavefront location $\Upsilon$ are given by Eq.~\eqref{eq:Shock_Amp_b0} and by the numerical integration of Eq.~\eqref{eq:shock_front_ODE}, respectively.}
	\label{fig:shock_num_varyk}
\end{figure}

Figures \ref{fig:aw_num} shows the evolution of an acceleration wave, while Figs.~\ref{fig:shock_num} and \ref{fig:shock_num_varyk} show the evolution of a shock wave in $U(X,T)$. The numerical solutions of the IBVP are compared to the theoretical results from singular surface theory discussed in Sect.~\ref{sec:periodic_SST}. The agreement at the wavefront $X=\Upsilon(T)$ is very good, however,  as $k$ becomes large (in Figs.~\ref{fig:aw_num} and \ref{fig:shock_num_varyk}), the wavefront becomes \emph{highly} localized, which leads to some small amount of numerical error, e.g., in Fig.~\ref{fig:shock_num_varyk}(f). The numerical solutions reveal many more features than the theoretical discussion. Specifically, we observe that the wave profile behind the wavefront, i.e., for $0<X<\Upsilon(T)$, is quite complex due to the periodic density profile, especially for the case of a shock wave in Fig.~\ref{fig:shock_num_varyk}(d,e,f). From Eq.~\eqref{eq:c_1}, we see that the periodic density profile necessitate a periodic sound speed. Thus, as the acceleration (or shock) wave propagates forward, acoustic disturbance emanate backwards from it as it has to slow down or speed up due to the variable sound speed. These disturbances reach the boundary  $X=0$, reflect and create this complex superposition that is particularly well illustrated in Fig.~\ref{fig:shock_num_varyk}.

\section{Closure}\label{sect:Closure}

Lastly,  we offer the following as possible, analytically tractable, extensions of the present study.

\begin{enumerate}
\item[$\bullet$] Consider vertically-running shock and acceleration waves under Taylor's~\cite{Taylor29} two-layer  atmosphere model; viz.:
\be
\varrho_{\rm a}(z)=
\varrho_{0}
\begin{cases}
\left(1- z/H_{1}\right)^{\frac{\gamma+1}{\gamma-1}}, &  z \in (0, z_{\rm i}), \\
\exp\left[-(z-z_{\rm i})/H_{2} \right]\left(1-z_{\rm i}/H_{1}\right)^{\frac{\gamma+1}{\gamma-1}}, & z \geq z_{\rm i},
\end{cases}
\en
with $\vartheta_{\rm a}(z)= \vartheta_{0}-\beta_{1}z/2$ for  $z\in (0, z_{\rm i})$ and $\vartheta_{\rm a}(z)=\vartheta_{\rm i}$ for  $z\geq z_{\rm i}$.  Here, $H_{1} :=2H_{\rm h}$ and $H_{2} :=\vartheta_{\rm i}/\beta$, the interface between the  layers lies at $z = z_{\rm i}$, and $\vartheta_{\rm i}=\vartheta_{0}-\beta_{1}z_{\rm i}/2$.

\item[$\bullet$]  The following outlines what is, perhaps, the most promising  approach  by which exact solutions to the simplest (i.e., \emph{full-Dirichlet}) version of IBVP~\eqref{eq:WE_IBVP_Signal2} might be derived:  Apply the Laplace transform to Eq.~\eqref{EoM:Periodic_Signal2} and the BCs, where the right-BC now reads $U(1,T)=0$, and then make use of the ICs to  get the BVP 
\begin{subequations}\label{BVP:b=zero}
\begin{align}
& \overline{U}_{XX}- s^2[1+\epsilon \cos(k\pi X)]\overline{U} =0, \quad X\in (0, 1),\\
& \overline{U}(0,s) = \overline{f}(s),\quad \overline{U}(1,s) = 0, \quad s>0,
\end{align}
\end{subequations}
the exact  solution of which is readily found to be
\begin{multline}\label{eq:Lap_Mathieu}
\overline{U}(X,s) = \frac{\overline{f}(s)}{ C\left(-\,\frac{4s^2}{k^2\pi^2}, \frac{2\epsilon s^2}{k^2\pi^2}, 0 \right)}  \Bigg\{ C\left(-\,\frac{4s^2}{k^2\pi^2}, \frac{2\epsilon s^2}{k^2\pi^2}, \tfrac{1}{2} k\pi X \right) \\
-\left[\frac{C\left(-\,\frac{4s^2}{k^2\pi^2}, \frac{2\epsilon s^2}{k^2\pi^2}, \tfrac{1}{2} k\pi \right)}{ S\left(-\,\frac{4s^2}{k^2\pi^2}, \frac{2\epsilon s^2}{k^2\pi^2}, \tfrac{1}{2} k\pi \right)}\right]\\
\times S\left( -\,\frac{4s^2}{k^2\pi^2}, \frac{2\epsilon s^2}{k^2\pi^2}, \tfrac{1}{2} k\pi X \right)  \Bigg\}.
\end{multline}
Here, $C(\varsigma_1, \varsigma_2, \zeta)$ and $S(\varsigma_1, \varsigma_2, \zeta)$ are the even and odd Mathieu functions~\cite{McL64}, respectively. 

In principle,  the exact time-domain solution, $U(X,T)$, can be determined by applying the `Inversion Theorem'~\cite{CJ63} (also known as the \emph{complex inversion formula}) to Eq.~\eqref{eq:Lap_Mathieu}. 

\item[$\bullet$] Examine signaling problems wherein the present linear equations of motion are replaced by their \emph{weakly-nonlinear}\footnote{As described in, e.g., Ref.~\cite{PMJ16}; i.e., the flow's Mach number (e.g., the ratio $W_0/c_0$ in Sect.~\ref{sect:Atmos}) is assumed to be small, but non-infinitesimal.} counterparts;  e.g., re-work the (weakly-nonlinear) IBVP analyzed in Ref.~\cite{QJMAM}, wherein  $f(t) \propto \sin(\omega t)$ was also used, assuming an inhomogeneous gas.

\end{enumerate}

\section*{Acknowledgments} 
The authors thank Profs.\ A.\ Rosato and M.\ Destrade for their kind invitation to contribute to this special issue, co-guest edited by M.D.\ and I.C.C., memorializing Prof.~G\'{e}rard Maugin. 
The authors also thank the  anonymous reviewer for his/her valuable comments and suggestions.
R.S.K.\ and P.M.J.\ were supported by ONR funding. I.C.C.\ thanks Prof.\ Kyle Mandli for helpful discussions on PyClaw.


\begin{thebibliography}{99}

\bibitem{Abr65}
M.\ Abramowitz, I.A.\ Stegun (Eds.), Handbook of Mathematical Functions, Dover, 1965.

\bibitem{BEM00}
A.\ Berezovski, J.\ Engelbrecht, G.A.\ Maugin, Thermoelastic wave propagation in inhomogeneous media, Arch.\ Appl.\ Mech.\ 70 (2000) 694--706, doi:10.1007/s004190000114.

\bibitem{BM02}
A.\ Berezovski, G.A.\ Maugin, Thermoelastic wave and front propagation, J.\ Thermal Stresses 25 (2002) 719--743, doi:10.1080/01495730290074504.

\bibitem{Bergmann46}
P.G.\ Bergmann, The wave equation in a medium with a variable index of refraction, J.\ Acoust.\ Soc.\ Am.\  17 (1946) 329--333, doi:10.1121/1.1916333.

\bibitem{Bland88} 
D.R.\ Bland,  Wave Theory and Applications, Oxford Univ.\ Press, 1988.

\bibitem{BH68}  
B.A.\ Boley,   R.B.\ Hetnarski, Propagation of discontinuities in coupled thermoelastic problems, J.\ Appl.\ Mech.\ (ASME) 35 (1968) 489--494, doi:10.1115/1.3601240.

\bibitem{Boyd12} 
J.P.\ Boyd,  Numerical, perturbative and Chebyshev inversion of the incomplete
elliptic integral of the second kind, Appl.\ Math.\ Comput.\ 218 (2012) 7005--7013, doi:10.1016/j.amc.2011.12.021.

\bibitem{CJ63} 
H.S.\ Carslaw, J.C.\ Jaeger, Operational Methods in Applied Mathematics, Dover, 1963.

\bibitem{chen1973} 
P.J.\ Chen,  Growth and decay of waves in solids, in: 
S.\ Fl\"ugge, C.\ Truesdell (Eds.), Handbuch der Physik, vol.~VIa/3, Springer, 1973, 
pp.~303--402. 

\bibitem{CC17}
I.C.\ Christov, C.I.\ Christov, On mechanical waves and Doppler shifts from moving boundaries, Math.\ Meth.\ Appl.\ Sci.\ 40 (2017) 4481--4492, doi:10.1002/mma.4318. 

\bibitem{QJMAM}
I.\ Christov, C.I.\ Christov, P.M.\ Jordan, Modeling weakly nonlinear acoustic wave propagation, {\it Q.\ J.\ Mech.\ Appl.\ Math.}\ 60 (2007) 473--495, doi:10.1093/qjmam/hbm017; {\it ibid.}\ 68 (2015) 231--233, doi:10.1093/qjmam/hbu023.

\bibitem{PyClaw1}
Clawpack Development Team, Clawpack software, \url{http://www.clawpack.org}, version~5.4.0, 2017.

\bibitem{C98}
D.G.\ Crighton, Propagation of finite-amplitude waves in fluids, in: M.J.\ Crocker (Ed.), Handbook of Acoustics, Wiley, 1998, chap.~17.


\bibitem{LeVeque99}
T.R.\ Fogarty, R.J.\ LeVeque, High-resolution finite-volume methods for acoustic waves
in periodic and random media, J.\ Acoust.\ Soc.\ Am.\  106 (1999) 17--28, doi:10.1121/1.428038.



\bibitem{PMJ16}
P.M.\ Jordan, A survey of weakly-nonlinear acoustic models: 1910--2009,  Mech.\ Res.\ Commun.\ 73 (2016) 127--139, doi:10.1016/j.mechrescom.2016.02.014.

\bibitem{PyClaw2}
D.I.\ Ketcheson et al., {PyClaw: Accessible, extensible, scalable tools for wave propagation problems}, SIAM J.\ Sci.\ Comput.\ 34 (2012) C210--C231, doi:10.1137/110856976.

\bibitem{Lamb06}
H.\ Lamb, Hydrodynamics, 3rd edn., Cambridge University Press, 1906.

\bibitem{Lamb09}
H.\ Lamb, On the theory of waves propagated vertically in the atmosphere, Proc.\ Lond.\ Math.\ Soc.\ (Ser.~2) 7 (1909) 122--141, doi: 10.1112/plms/s2-7.1.122.

\bibitem{Lamb11}
H.\ Lamb, On atmospheric oscillations, Proc.\ R.\ Soc.\ Lond.\ A 84 (1911) 551--572, doi:
10.1098/rspa.1911.0008.

\bibitem{Lamb45}
H.\ Lamb, Hydrodynamics, 6th edn., Dover, 1945.

\bibitem{Lanczos}
C.\ Lanczos, Applied Analysis, Prentice Hall, 1956, pp.~221--227.

\bibitem{L97}
R.J.\ LeVeque, Wave propagation algorithms for multidimensional hyperbolic systems, J.\ Comput.\ Phys.\ 131 (1997) 327--353, doi:10.1006/jcph.1996.5603.

\bibitem{LeVeque02}
R.J.\ LeVeque, Finite Volume Methods for Hyperbolic Problems, Cambridge University Press, 2002.

\bibitem{M93}
G.A.\ Maugin, Material Inhomogeneities in Elasticity,  Chapman \& Hall, 1993, chap.~4.

\bibitem{M98}
G.A.\ Maugin, On shock waves and phase-transition fronts in continua, ARI 50 (1998) 141--150, doi:10.1007/s007770050008

\bibitem{McL64}
N.W.\ McLachlan, The Theory and Application of Mathieu Functions, Dover, 1964.

\bibitem{OW}
V.E.\ Ostashev, D.K.\ Wilson, Acoustics in Moving Inhomogeneous Media, 2nd edn., CRC Press, 2015.

\bibitem{P89}
A.D.\ Pierce, Acoustics: An Introduction to its Physical Principles and Applications, Acoustical Society of America, 1989.

\bibitem{Rayleigh}
Lord Rayleigh, On the vibrations of an atmosphere, Phil.\ Mag.\ (Ser.~5) 29 (1890) 173--180, doi:10.1080/14786449008619921.

\bibitem{S11}
B.\ Straughan, Heat Waves, in: Applied Mathematical Sciences, vol.~177, Springer, 2011, chap.~4.

\bibitem{Taylor29}
G.I.\ Taylor, Waves and tides in the atmosphere, Proc.\ R.\ Soc.\ Lond.\ A 126 (1929) 169--183, doi: 10.1098/rspa.1929.0213.

\bibitem{T72}
P.A.\ Thompson, Compressible-Fluid Dynamics, McGraw--Hill, 1972.

\bibitem{tt60}
C.\ Truesdell, R.A.\ Toupin,  The classical field theories, 
in: S.\ Fl\"{u}gge (Ed.), Handbuch der Physik, vol.\ III/1, Springer, 1960, pp.~491--529.

\bibitem{Tzou97}
D.Y.\ Tzou, Macro- to Microscale Heat Transfer: The Lagging Behavior, Taylor \& Francis, 1997, sect.~2.5.1.

\bibitem{Walsh69}
E.K.\ Walsh, Development of shock waves in atmospheres with density and temperature variations, Phys.\ Fluids 12 (1969) 757--763.

\bibitem{EWW}
E.W.\   Weisstein, Kepler's Equation. From MathWorld---A Wolfram Web Resource (http://mathworld.wolfram.com/KeplersEquation.html).

\bibitem{W74}
G.B.\ Whitham, Linear and Nonlinear Waves, Wiley,  1974. 

\bibitem{WPO}
D.K.\ Wilson, C.L.\ Pettit, V.E.\ Ostashev, Sound propagation in the atmospheric boundary layer,  Acoust.\ Today 11(2) (2015) 44--53.


\end{thebibliography}
\end{document}